\documentclass[12pt]{cernart}

\usepackage{epsfig}
\usepackage{graphicx} 
\usepackage{amsmath,amssymb}
\usepackage{lineno}

\newcommand{\D}[2]{\frac{\partial #2}{\partial #1}}

\newcommand{\defmath}[2] {\def#1{\ifmmode{#2}\else\mbox{${#2}$}\fi}}
\newcommand{\defdecay}[2] {\def#1{\ifmmode{#2}\else\mbox{${#2}$}\fi}}

\newcommand{\Kpienu} {\mbox{$K^\pm \rightarrow\pi^0 e^\pm\nu$}}
\newcommand{\Kpimunu}{\mbox{$K^\pm \rightarrow\pi^0 \mu^\pm\nu$}}
\newcommand{\Kpipi}   {\mbox{$K^\pm \rightarrow \pi^\pm \pi^0$}}

\newcommand{\rkekp} {\mbox{${\cal R}_{K e 3 / K2\pi}$}}
\newcommand{\rkmukp} {\mbox{${\cal R}_{K \mu 3 / K2\pi}$}}
\newcommand{\rkmuke} {\mbox{${\cal R}_{K \mu 3 / Ke3}$}}

\newcommand{\defunit}[2] {\def#1{\ifmmode\mathrm{#2}\else\mbox{$\mathrm{#2}$}\fi
}} 

\defunit{\mum}{\mu m}
\defunit{\mus}{\mu s}
\defunit{\taus}{\tau_S}
\defunit{\degrees}{^{\circ}}
\defunit{\xo}{X_0}

\defdecay{\k} {K}
\defdecay{\kz} {K^0}
\defdecay{\kbar} {\overline{K^{0}}}
\defdecay{\cascadebar} {\overline{\Xi^0}}
\defdecay{\lambdabar} {\overline{\Lambda}}
\defdecay{\kone} {K_1}
\defdecay{\ktwo} {K_2}
\defdecay{\ks} {K_S}
\defdecay{\kl} {K_L}
\defdecay{\ksl} {K_{S,L}}
\defdecay{\KL} {\kl}
\defdecay{\KS} {\ks}
\defdecay{\sbar} {\overline{\mathrm{s}}}
\defdecay{\dbar} {\overline{\mathrm{d}}}

\defdecay{\pipi} {\pi^{+}\pi^{-}}
\defdecay{\twopi} {2\pi}
\defdecay{\kpipi} {\k \rightarrow \pipi}
\defdecay{\ktwopi} {\ks \rightarrow \twopi}
\defdecay{\kspipi} {\ks \rightarrow \pipi}
\defdecay{\kstwopi} {\kl \rightarrow \twopi}
\defdecay{\klpipi} {\kl \rightarrow \pipi}
\defdecay{\kltwopi} {\k \rightarrow \twopi}
\defdecay{\pipipi} {\pi\pi\pi}
\defdecay{\threepi} {3\pi}

\newcommand{\vus}{\mbox{$V_{us}$}}

\defdecay{\pz}{\pi^0}
\defdecay{\pio} {\pi^0}
\defdecay{\twopio} {2 \pi^0}
\defdecay{\piopio} {\pi^0 \pi^0}
\defdecay{\pipin} {\piopio}
\defdecay{\ktwopio} {\k \rightarrow \twopio}
\defdecay{\kpiopio} {\k \rightarrow \piopio}
\defdecay{\kltwopio} {\kl \rightarrow \twopio}
\defdecay{\klpiopio} {\kl \rightarrow \piopio}
\defdecay{\kstwopio} {\ks \rightarrow \twopio}
\defdecay{\kspiopio} {\ks \rightarrow \piopio}
\defdecay{\klthreepich} {\kl \rightarrow \pio\pi^{+}\pi^{-}}

\defdecay{\kspimumu} {\ks \rightarrow \pio \mu^+ \mu^-}
\defdecay{\klpimumu} {\kl \rightarrow \pio \mu^+ \mu^-}

\defdecay{\kspiee} {\ks \rightarrow \pio e^+ e^-}
\defdecay{\klpiee} {\kl \rightarrow \pio e^+ e^-}
\defdecay{\kspipiD} {\ks \rightarrow \pio \pioD }
\defdecay{\kspipid} {\ks \rightarrow \pio \piod }
\defdecay{\kspiDpiD} {\ks \rightarrow \pioD \pioD }
\defdecay{\kspidpid} {\ks \rightarrow \piod \piod }
\defdecay{\kspipidd} {\ks \rightarrow \pio \piodd}
\defdecay{\kpiee} {K \rightarrow \pi e e}
\defdecay{\piee} {\pio e^+ e^-}
\defdecay{\pipid} {\pio \piod}
\defdecay{\pidpid} {\piod \piod}
\defdecay{\pipidd} {\pio \piodd}
\defdecay{\p0pipi} {\pio \pi^{+} \pi^{-} }
\defdecay{\kleegg} {\kl\rightarrow e e\gamma\gamma}
\defdecay{\klmumugg} {\kl\rightarrow \mu\mu\gamma\gamma}
\defdecay{\klpigg} {\kl\rightarrow \pio \gamma \gamma}
\defdecay{\kseegg} {\ks\rightarrow e e\gamma\gamma}
\defdecay{\eegg} {e e\gamma\gamma}
\defdecay{\kleeg} {\kl\rightarrow e e\gamma}
\defdecay{\klp0pipi} {\kl\rightarrow \pio \pi^+ \pi^- }
\defdecay{\BRkspiee} {B ( \ks \rightarrow \pio e^+ e^- ) }
\defdecay{\BRklpiee} {B ( \kl \rightarrow \pio e^+ e^- ) }
\defdecay{\BRkspipid} {B ( \kspipid ) }
\defdecay{\BRkspidpid} {B ( \kspidpid ) }
\defdecay{\BRklp0pipi} {B ( \klp0pipi) }
\defdecay{\BRp0pipi} {B ( \p0pipi) }
\defdecay{\Xilampi} {\Xi^0 \rightarrow \Lambda \pio}
\defdecay{\lamppi} {\Lambda \rightarrow p \pi^{-}}
\defdecay{\Xilamppipio} {\Xi^{0}\rightarrow\Lambda (p \pi^{-})\pio }
\defdecay{\lampev} {\Lambda \rightarrow p e^{-} \nu}
\defdecay{\Xilampevpio} {\Xi^{0}\rightarrow\Lambda (p e^{-} \nu)\pio }
\defdecay{\Xisigmaenv} {\Xi^0 \rightarrow \Sigma^{+} e^{-} \nu}
\defdecay{\sigppio} {\Sigma^{+} \rightarrow  p \pio}
\defdecay{\Xisigmappioenv} {\Xi^0 \rightarrow \Sigma^{+}(p \pio) e^{-} \nu}
\defdecay{\AXilampi} {\overline{\Xi^0}\rightarrow \overline{\Lambda} \pio  }
\defdecay{\Alamppi} {\overline{\Lambda} \rightarrow \overline{p} \pi^{+}}
\defdecay{\Alampev} {\overline{\Lambda} \rightarrow \overline{p} e^{+} \, \overline{\nu}}
\defdecay{\AXisigmaenv} {\overline{\Xi^0} \rightarrow \overline{\Sigma^{+}} e^{+} \nu}
\defdecay{\Asigppio} {\overline{\Sigma^{+}} \rightarrow  \overline{p} \pio}
\defdecay{\klkef} {\kl\rightarrow \pio \pi^{\pm} e^{\mp} \nu }
\defdecay{\kef} {\pio \pi^{\pm} e^{\mp} \nu }
\defdecay{\klketh} {\kl\rightarrow \pi^{\pm} e^{\mp} \nu }
\defdecay{\keth} {\pi^{\pm} e^{\mp} \nu }

\defdecay{\dalconv} {{\bf\boldmath \pio}(\gamma\gamma){\bf\boldmath \piod}(\gamma_{conv}(ee\hspace {-3.5mm}\nearrow)ee\hspace {-3.5mm}\nearrow)}
\defdecay{\ndalconv} {{\bf\boldmath \pio}(\gamma_{conv}(ee\hspace {-3.5mm}\nearrow)\gamma){\bf\boldmath \piod}(\gamma ee\hspace {-3.5mm}\nearrow)}
\defdecay{\dalcomp} {{\bf\boldmath \pio}(\gamma\gamma){\bf \boldmath\piod}(\gamma_{comp}(e^{-})e^{+}e^{-}\hspace {-1.5mm}\nearrow)}

\defdecay{\pipiconv} {{\bf\boldmath \pio}(\gamma\gamma){\bf\boldmath \pio}(\gamma_{conv}(ee\hspace {-6.5mm}\nearrow)\gamma_{conv}(ee\hspace {-6.5mm}\nearrow))}

\defdecay{\klpipipic} {\kl \rightarrow \pio \pi^+ \pi^-}
\defdecay{\kspipic} {\ks \rightarrow \pi^+ \pi^-}
\defdecay{\kspiee} {\ks \rightarrow \pio e^+ e^-}
\defdecay{\kspimumu} {\ks \rightarrow \pio \mu^+ \mu^-}
\defdecay{\klmumugg} {\kl \rightarrow \mu^+ \mu^- \gamma \gamma}
\defdecay{\klpiogg} {\kl \rightarrow \pio \gamma \gamma}
\defdecay{\kspipig} {\ks \rightarrow \pi^+ \pi^- \gamma}
\defdecay{\klmufour} {\kl \rightarrow \pio \pi^{\pm} \mu^{\mp} \nu}
\defdecay{\klpimunug} {\kl \rightarrow \pi^{\pm} \mu^{\mp} \nu \gamma}

\defdecay{\pip} {\pi^+}
\defdecay{\pim} {\pi^-}
\defdecay{\twopic} {\pi^+ \pi^-}
\defdecay{\pipic} {\twopic}
\defdecay{\ktwopic} {\k \rightarrow \twopic}
\defdecay{\kltwopic} {\kl \rightarrow \twopic}
\defdecay{\kstwopic} {\ks \rightarrow \twopic}

\defdecay{\piod} {\pi_{D}^0}
\defdecay{\pioD} {\pi_{Dalitz}^0}
\defdecay{\piodd} {\pi_{DD}^0}
\defdecay{\piopiod} {\pi^0 \pi_{D}^0}
\defdecay{\eeg} {ee\gamma}
\defdecay{\pioeeg} {\pi^0 \rightarrow \eeg}
\defdecay{\kpiopiod} {\k \rightarrow \piopiod}

\defdecay{\kethree} {\mathrm{K_{e3}}}
\defdecay{\pienu} {\pi e \nu}
\defdecay{\klethree} {\kl \rightarrow \pienu}
\defdecay{\kmuthree} {\mathrm{K_{\mu 3}}}
\defdecay{\pimunu} {\pi \mu \nu}
\defdecay{\klmuthree} {\kl \rightarrow \pimunu}
\defdecay{\threepio} {3\pi^0}
\defdecay{\klthreepio} {\kl \rightarrow \threepio}

\defdecay{\Lam}{\Lambda}
\defdecay{\Lambar}{\bar{\Lambda}}
\defdecay{\etagg}{\eta \rightarrow \gamma \gamma}
\defdecay{\etathreepio} {\eta \rightarrow \threepio}
\defdecay{\piogg} {\pi^0 \rightarrow \gamma \gamma}
\defdecay{\meegg} {m_{ee \gamma \gamma}}
\defdecay{\meeggg} {m_{ee \gamma \gamma \gamma}}
\defdecay{\mgg} {m_{\gamma \gamma}}
\defdecay{\mmumugg} {m_{\mu\mu \gamma \gamma}}

\defmath{\dvertex}{d_{vertex}}
\defmath{\mgg}{m_{\gamma\gamma}}
\defmath{\chisq}{\chi^2}
\defmath{\rel}{\chi^2}
\defmath{\mone}{m_1}
\defmath{\mtwo}{m_2}

\defmath{\pk}{p_K}
\defmath{\pt}{p_T}
\defmath{\ptp}{{p_T}'}
\defmath{\ptpsq}{{p'_T}^2}
\defmath{\mpp}{m_{\pi\pi}}

\defmath{\asl} {\alpha_{SL}}
\defmath{\asloo} {\alpha^{00}_{SL}}
\defmath{\aslpm} {\alpha^{+-}_{SL}}
\defmath{\Dasl} {\Delta \alpha_{SL}}

\defmath{\als} {\alpha_{LS}}
\defmath{\alsoo} {\alpha^{00}_{LS}}
\defmath{\alspm} {\alpha^{+-}_{LS}}
\defmath{\Dals} {\Delta \alpha_{LS}}

\defmath{\btag} {\beta_{tag}}
\defmath{\btagoo} {\beta^{00}_{tag}}
\defmath{\btagpm} {\beta^{+-}_{tag}}
\defmath{\Dbtag} {\Delta \beta_{tag}}

\defmath{\wpm} {W^{+-}}
\defmath{\woo} {W^{00}}
\defmath{\wooo} {W^{000}}
\defmath{\Dw} {\Delta W}

\defmath{\wt}{W(\tau)}
\defmath{\Dp}{\mathrm{D_p}}

\defmath{\etas}{\eta_S}
\defmath{\etal}{\eta_L}
\defmath{\etasl}{\eta_{S,L}}
\defmath{\lamc}{\lambda^{+-}}
\defmath{\lamn}{\lambda^{00}}
\defmath{\DRint}{(\DR)_{\mbox{\scriptsize intensity}}}
\defmath{\DRgeom}{(\DR)_{\mbox{\scriptsize geometry}}}

\defmath{\R}{R}
\defmath{\DR}{\Delta \R}
\defmath{\epp}{\varepsilon^{\prime}}
\defmath{\vep}{\varepsilon}
\defmath{\epe}{\epp/\vep}
\defmath{\Ree}{\mathcal{R}\!\mathit{e}(\eprime/\epsilon)}
\defmath{\epm}{\eta_{+-}}
\defmath{\eoo}{\eta_{00}}

\defmath{\mum}{\mu\mathrm{m}}
\defmath{\mus}{\mu\mathrm{s}}
\defmath{\degrees}{^{\circ}}
\defmath{\taus}{\tau_S}
\defmath{\taul}{\tau_L}

\defmath{\about}{\sim}
\defmath{\eop}{E/p}
\defmath{\Qx} {Q_x}
\defmath{\twotrack}{2track}
\defmath{\etot}{E_{tot}}
\defmath{\mk}{m_K}
\defmath{\stat}{\mbox{stat}}
\defmath{\syst}{\mbox{syst}}
\defmath{\rcog}{R_{cog}}
\defmath{\mm}{m}
\defmath{\mee}{m_{ee}}
\defmath{\mpi}{m_{\pio}}

\defdecay{\pimumu} {\pio \mu^+ \mu^-}

\usepackage{amssymb}

\begin{document}
\begin{titlepage}
\docnum{CERN--EP/2006-039}
\date{Dec. 8 2006}
\title{\bf \Large Measurements of Charged Kaon Semileptonic Decay  
Branching Fractions  \Kpimunu\ and \Kpienu\
and Their Ratio }
\begin{center}
\  \\
J.R.~Batley,
C.~Lazzeroni,
D.J.~Munday,
M.W.~Slater,
S.A.~Wotton \\
{\em \small Cavendish Laboratory, University of Cambridge, Cambridge, CB3 0HE,
U.K.$\,$\footnotemark[1]} \\[0.2cm]
R.~Arcidiacono$\,$\footnotemark[2],
G.~Bocquet,
N.~Cabibbo,
A.~Ceccucci,
D.~Cundy$\,$\footnotemark[3],
V.~Falaleev,
M.~Fidecaro,
L.~Gatignon,
A.~Gonidec,
W.~Kubischta,
A.~Norton$\,$\footnotemark[4],
M.~Patel,
A.~Peters \\
{\em \small CERN, CH-1211 Geneva 23, Switzerland} \\[0.2cm]
S.~Balev,
P.L.~Frabetti,
E.~Goudzovski,
P.~Hristov$\,$\footnotemark[5],
V.~Kekelidze$\,$\footnotemark[5],
V.~Kozhuharov,
L.~Litov,
D.~Madigozhin,
E.~Marinova,
N.~Molokanova,
I.~Polenkevich,
Yu.~Potrebenikov,
S.~Stoynev$\,$\footnotemark[6],
A.~Zinchenko \\
{\em \small Joint Institute for Nuclear Research, Dubna, Russian    Federation}
\\[0.2cm]
E.~Monnier$\,$\footnotemark[7],
E.~Swallow,
R.~Winston\\
{\em \small The Enrico Fermi Institute, The University of Chicago, Chicago, IL 60126,
U.S.A.}\\[0.2cm]
 P.~Rubin,
 A.~Walker \\
{\em \small Department of Physics and Astronomy, University of
 Edinburgh, \\
JCMB King's Buildings, Mayfield Road, Edinburgh,    EH9 3JZ, U.K.} \\[0.2cm]
W.~Baldini,
A.~Cotta Ramusino,
P.~Dalpiaz,
C.~Damiani,
M.~Fiorini,
A.~Gianoli,
M.~Martini,
F.~Petrucci,
M.~Savri\'e,
M.~Scarpa,
H.~Wahl \\
{\em \small Dipartimento di Fisica dell'Universit\`a e Sezione    dell'INFN di Ferrara,
I-44100 Ferrara, Italy} \\[0.2cm]
A.~Bizzeti$\,$\footnotemark[8],
M.~Calvetti,
E.~Celeghini,
E.~Iacopini,
M.~Lenti,
F.~Martelli$\,$\footnotemark[9], \\
G.~Ruggiero$\,$\footnotemark[5],
M.~Veltri$\,$\footnotemark[9] \\
{\em \small Dipartimento di Fisica dell'Universit\`a e Sezione    dell'INFN di Firenze,
I-50125 Firenze, Italy} \\[0.2cm]
M.~Behler,
K.~Eppard,
K.~Kleinknecht,
P.~Marouelli,
L.~Masetti,
U.~Moosbrugger,\\
C.~Morales Morales,
B.~Renk,
M.~Wache,
R.~Wanke,
A.~Winhart \\
{\em \small Institut f\"ur Physik, Universit\"at Mainz, D-55099 Mainz,
Germany$\,$\footnotemark[10]} \\[0.2cm]
D.~Coward$\,$\footnotemark[11],
A.~Dabrowski,
T.~Fonseca Martin$\,$\footnotemark[5],
M.~Shieh,
M.~Szleper,
M.~Velasco,
M.D.~Wood$\,$\footnotemark[12] \\
{\em \small Department of Physics and Astronomy,
Northwestern University, \\ Evanston, IL 60208-3112, U.S.A.}
 \\[0.2cm]
G.~Anzivino,
P.~Cenci,
E.~Imbergamo,
M.~Pepe,
M.C.~Petrucci,
M.~Piccini$\,$\footnotemark[5],
M.~Raggi,
M.~Valdata-Nappi \\
{\em \small Dipartimento di Fisica dell'Universit\`a e Sezione    dell'INFN di Perugia,
I-06100 Perugia, Italy} \\[0.2cm]
C.~Cerri,
G.~Collazuol,
F.~Costantini,
L.~DiLella,
N.~Doble,
R.~Fantechi,
L.~Fiorini$\,$\footnotemark[13],
S.~Giudici,
G.~Lamanna,
I.~Mannelli,
A.~Michetti,
G.~Pierazzini,
M.~Sozzi \\
{\em \small Dipartimento di Fisica, Scuola Normale Superiore e Sezione
dell'INFN di Pisa, \\
I-56100 Pisa, Italy} \\[0.2cm]
B.~Bloch-Devaux,
C.~Cheshkov$\,$\footnotemark[5],
J.B.~Ch\`eze,
M.~De Beer,
J.~Derr\'e,
G.~Marel,
E.~Mazzucato,
B.~Peyaud,
B.~Vallage \\
{\em \small DSM/DAPNIA - CEA Saclay, F-91191 Gif-sur-Yvette, France} \\[0.2cm]
M.~Holder,
A.~Maier$\,$\footnotemark[5],
M.~Ziolkowski \\
{\em \small Fachbereich Physik, Universit\"at Siegen, D-57068 Siegen,
Germany$\,$\footnotemark[14]} \\[0.2cm]
S.~Bifani,
C.~Biino,
N.~Cartiglia,
M.~Clemencic$\,$\footnotemark[5],
S.~Goy Lopez,
F.~Marchetto \\
{\em \small Dipartimento di Fisica Sperimentale dell'Universit\`a e
Sezione dell'INFN di Torino, \\
I-10125 Torino, Italy} \\[0.2cm]
H.~Dibon,
M.~Jeitler,
M.~Markytan,
I.~Mikulec,
G.~Neuhofer,
L.~Widhalm \\
{\em \small \"Osterreichische Akademie der Wissenschaften, Institut  f\"ur
Hochenergiephysik, \\
A-10560 Wien, Austria$\,$\footnotemark[15]} \\[1.5cm]
\rm
\setcounter{footnote}{0}
\footnotetext[1]{Funded by the U.K.    Particle Physics and Astronomy Research Council}
\footnotetext[2]{Present address: Dipartimento di Fisica Sperimentale dell'Universit\
Sezione dell'INFN di Torino, \\
I-10125 Torino, Italy }
\footnotetext[3]{Present address: Istituto di Cosmogeofisica del CNR
di Torino, I-10133 Torino, Italy}
\footnotetext[4]{Present address: Dipartimento di Fisica dell'Universit\`a e Sezione
Ferrara, I-44100 Ferrara, Italy}
\footnotetext[5]{Present address: CERN, CH-1211 Geneva 23, Switzerland}
\footnotetext[6]{Present address: Department of Physics and Astronomy,
Northwestern University,  Evanston, IL 60208-3112, U.S.A.}
\footnotetext[7]{Also at Centre de Physique des Particules de Marseille, IN2P3-CNRS,
Universit\'e
de la M\'editerran\'ee, Marseille, France}
\footnotetext[8] {Also Dipartimento di Fisica, Universit\`a di Modena, I-41100  Modena,
Italy}
\footnotetext[9]{Istituto di Fisica, Universit\`a di Urbino, I-61029  Urbino, Italy}
\footnotetext[10]{Funded by the German Federal Minister for Education
and research under contract 05HK1UM1/1}
\footnotetext[11]{Permanent address: SLAC, Stanford University, Menlo Park, CA 94025,
U.S.A.}
\footnotetext[12]{Present address: UCLA, Los Angeles, CA 90024, U.S.A.}
\footnotetext[13]{Present address: Institut de Fisica d~RAltes Energies, Universitat
A Barcelona, E-08193 Bellaterra (Barcelona), Spain}
\footnotetext[14]{Funded by the German Federal Minister for Research and Technology
(BMBF) under contract 056SI74}
\footnotetext[15]{Funded by the Austrian Ministry for Traffic
and Research under the contract GZ 616.360/2-IV GZ 616.363/2-VIII,
and by the Fonds f\"ur Wissenschaft und Forschung FWF Nr.~P08929-PHY}
\end{center}
\begin{center}
{\it Published in European Physical Journal C {\bf 50} 2 (2007), contains erratum submitted to European Physical Journal C}
\end{center}
\end{titlepage}

\setcounter{footnote}{0}
\begin{abstract}
Measured ratios of decay rates for $\rkekp$, $\rkmukp$ and  $\rkmuke$
 are presented.
These measurements are based on $K^\pm$ decays collected in a dedicated run in
2003 by the NA48/2 experiment at CERN. The results obtained are
$\rkekp  = 0.2470\pm 0.0009 (stat)\pm 0.0004 (syst)$ and
$\rkmukp = 0.1637\pm 0.0006 (stat)\pm 0.0003 (syst)$.
Using the PDG average for
the $K^\pm \rightarrow\pi^\pm \pi^0$ normalisation mode, both values are
found to be larger
than the current values given by the Particle Data Book and lead to a
larger magnitude of the $|V_{us}|$ CKM element than previously
accepted.  When combined with the latest Particle Data Book value of $|V_{ud}|$, the result is
 in agreement with unitarity of the CKM matrix.
In addition,
a new measured value of $\rkmuke = 0.663\pm 0.003(stat)\pm 0.001(syst)$ is compared to
the
semi-empirical predictions based on the latest form factor measurements.
\end{abstract}

\vspace{0.2cm}

\section{Introduction}
\label{sec:Introduction}
New measurements of the charged kaon semileptonic decays,
$K^{\pm}\rightarrow \pi^{0} e^{\pm} \nu$ ($K_{e3}$)
and $K^{\pm}\rightarrow \pi^{0} \mu^{\pm} \nu$ ($K_{\mu3}$), are presented.
These measurements are based on $K^\pm$ decays collected
in 2003 by the NA48/2 Collaboration at the CERN SPS.

The measured ratios are:
%
\begin{equation}
   \rkekp  \equiv
      \frac{ \Gamma(\Kpienu) }
           { \Gamma(\Kpipi) },~~~~~~~~~~~
   \rkmukp  \equiv
      \frac{ \Gamma(\Kpimunu) }
           { \Gamma(\Kpipi) }
   \label{eq:rkpilnurad}
\end{equation}
and
\begin{equation}
   \rkmuke  \equiv
      \frac{ \Gamma(\Kpimunu) }{ \Gamma(\Kpienu) }.
\end{equation}
In both  the numerator and denominator, the decays
contain   a charged track and  at least two photons originating from a
$\pi^0$ decay, thus leading to a partial cancellation in the acceptance and
reconstruction uncertainties.
Contributions from internal bremsstrahlung are included
for all three decay modes.
The general reconstruction methods are the same for all three
measurements but the event selection varies because different particle
identification criteria are applied.

The main interest in measuring  these quantities is
to extract: (1)~the individual semileptonic decay
widths needed to determine the $V_{us}$ element in the
Cabibbo-Kobayashi-Maskawa\,(CKM) quark mixing matrix; and
(2)~the ratio $\Gamma(K_{\mu 3})/\Gamma(K_{e3})$  which is
a function of the slope parameters of the form factors.
On the assumption of $\mu-e$ universality this ratio
provides a consistency check between measurements of the
 form factors and of  the partial decay widths.

This paper is organized as follows. Section~\ref{sec:phenom}
outlines  the phenomenological description of semileptonic decays.
The experimental setup is described in Section~\ref{sec:apparatus}.
The event selection and all corrections applied to the
data are presented in Section~\ref{sec:anal}.  Section~\ref{sec:extraction} includes
a detailed description of the acceptance corrections and radiative effects.
Finally, the results are presented in Section~\ref{sec:results}, and their
comparison with theory and their impact on $V_{us}$ are given in
Section~\ref{sec:dis}.

\section{Decay rates for semileptonic decays}
\label{sec:phenom}
The decay rate for charged semileptonic decays can be written as
follows~\cite{bib:Leu_Roos}:
\begin{equation}
\Gamma(K_{\ell 3}) = { G_F^2 \over 384 \pi^3} m_K^5 S_{EW} |V_{us}|^2
|f_+(0)|^2 I_K^\ell (1+\delta_K^\ell) \; \;,
\label{eq:vus}
\end{equation}
where $\ell$ refers to either $e$ or $\mu$, $G_F $ is the Fermi constant,
$m_K$ is the kaon mass, $S_{EW}$ is the  short-distance radiative correction,
$(1+\delta_K)\simeq (1+\delta_{SU(2)}^\ell +\delta_{EM}^\ell)^2$ is the
model-dependent long-distance
correction with  contributions due to isospin breaking in strong\,($SU(2)$)
 and
electromagnetic\,($EM$)  interactions, $f_+(0)$ is the
form factor at zero momentum transfer ($t=0$) for the
$\ell \nu$ system. The remaining term,
$I_K^\ell$, is the result of the phase space integration
after factoring out $f_+(0)$, and is defined as~\cite{bib:Leu_Roos}:
\begin{eqnarray}
I_K^\ell &=& \frac{1}{m_K^8}\int_{m_\ell^2}^{(m_K-m_{\pi})^2}\frac{dt}{2t^3}
(t-m_\ell^2)^2T^{1/2}(t,m_K^2,m_{\pi}^2) \nonumber \\
&\times&\left\{ T(t,m_K^2,m_{\pi}^2)(2t+m_\ell^2)|{f}_+(t)/{f}_+(0)|^2 +
3m_\ell^2(m_K^2-m_{\pi}^2)^2|{f}_0(t)/{f}_+(0)|^2\right\},
\label{eq:int}
\end{eqnarray}
where
$T(x,y,z)\equiv x^2+y^2+z^2-2xy-2xz-2yz$, and $m_{\pi}$ and $m_{\ell}$
are the masses of the $\pi^0$ and the charged lepton, respectively.
The two form factors correspond to the
angular momentum configuration of the $K-\pi$ system, with
$f_+ (t)$ representing the vector form factor, while $f_0 (t)$ is the
scalar form factor.
The $t$ dependence of the form factors  can be   described using a
quadratic\,(linear)
approximation for the vector\,(scalar) term:
\begin{equation}
\begin{array}{l}
{f}_+(t) = f_+(0)\left(1 + \lambda'_+ \frac{\textstyle t}{\textstyle  m_{\pi^\pm}^2} +
\frac{\textstyle 1}{\textstyle 2} \lambda''_+
 \frac{\textstyle t^2}{\textstyle  m_{\pi^\pm}^4}\right)~~{\rm and}~~
f_{0} (t) = f_+(0)\left(1+\lambda_{0} \frac{\textstyle t}{\textstyle m_{\pi^\pm}^2}
\right) \; ,
\\
\label{eq:q2}
\end{array}
\end{equation}
where $f_+(0)$ is obtained from theory, and $\lambda'_+$, $\lambda''_+$ and $\lambda_0$
 are measured~\cite{bib:PDG}.
The second
term in $I_K^\ell$ shows that $K_{\mu3}$ decays are more sensitive to
the scalar form factor than $K_{e3}$.

Other ways to describe the momentum dependence of the form factors
are also  considered and discussed in Section~\ref{sec:extraction}.

Assuming $\mu-e$ universality, the $\Gamma(K_{\mu3})/\Gamma(K_{e3})$ ratio
is predicted to be\,\cite{bib:9411311}:
\begin{equation}
\label{eq:ratio}
\rkmuke \equiv \Gamma (K_{\mu3}) / \Gamma (K_{e3}) =
\frac{
0.645 +2.087 \lambda_+ +1.464 \lambda_0 +3.375 \lambda_+^2 +2.573 \lambda_0^2}
{1 +3.457 \lambda_+ +4.783 \lambda_+^2}.
\end{equation}
This semi-empirical formula assumes a linear approximation
for the form factors,
$f_{+,0} (t) = f_+ (0)(1+\lambda_{+,0} {t \over m_{\pi^\pm}^2} )$.

If the assumption of $\mu-e$  universality is removed, Eq.\,(\ref{eq:vus}) implies
that:
\begin{eqnarray}
\label{eq:no_lepton1}
\rkmuke &=&[g_{\mu}f_{+}^{\mu}(0)/g_{e}f_{+}^e(0)]^2 \times (I_K^\mu  (1+\delta_K^\mu
))/ (I_K^e  (1+\delta_K^e)).
\end{eqnarray}
where $g$ is the weak coupling constant for the lepton current.
The $(1+\delta_K^e)/(1+\delta_K^\mu )$ ratio for charged kaons
 is very close to unity because the
$\delta_K^\ell$'s are  dominated by the few percent $SU(2)$ correction that
is in common between the electron and the muon channel. The $EM$
corrections are  at the per mille level and compatible with zero within errors
\cite{bib:Cirigliano,bib:gin,bib:wgr}.  This is not necessarily true for neutral kaons
because there is no  $SU(2)$ correction and also the electromagnetic correction
for the muons is larger.

\section{Experimental setup}
\label{sec:apparatus}

\subsection{Beam}
The experiment uses
simultaneous $K^+$ and $K^-$ beams                 produced by 400\,GeV
protons impinging on a Be target. The beam has particles of opposite
charge with a central
momentum of 60\,GeV/c and a momentum band of $\pm 3.8$\% produced at zero angle.
Both beams are selected by a system of dipole magnets forming an  achromat,
along with
focusing quadrupoles, muon sweepers and collimators.  The spill length is
about        4.5\,s out of a 16.8\,s cycle time, and the proton intensity
is fairly constant during the spill with a mean of $5 \times 10^{10}$
protons per spill.  The positive (negative)
kaon flux at the entrance of the decay volume is $3.2\times 10^6 (1.8\times
10^6)$ particles per spill.

For the measurements presented here, the proton beam intensity is
reduced  from its nominal value so that the data-acquisition system can
handle the rate of the  minimum bias trigger.
The  $K^+/K^-\simeq 1.78$ flux ratio is given by their production rate at the
Be-target.

\subsection{Detector}

The final beam
collimator is immediately followed by a 114\,m long cylindrical tank, which
is evacuated
in order to minimize the interactions of the decay products.
The tank
is terminated by a 0.3\% radiation length\,($X_0$) thick Kevlar window,
except in a region close to the beam which continues in a vacuum pipe through
the centre of the downstream detectors.  The detector is located downstream
of this Kevlar window.


The tracking is performed with a spectrometer housed in a helium gas volume.
It  consists of two drift chambers before and two after a dipole magnet with
a horizontal transverse momentum kick of 120\,MeV/c. Each chamber has four
views, each of which has two sense wire planes.  The resulting space points
are  typically reconstructed with a resolution of 150\,$\mu$m in each
projection.  The momentum resolution of the spectrometer is
$\sigma_p/p = 1.02 \% \oplus 0.044\% \cdot p$,
 where $p$ is in GeV/c.
The track time resolution is 1.4\,ns.

The detection and measurement of electromagnetic showers are  performed with
a 27\,$X_0$-deep liquid krypton calorimeter (LKr) that has an  energy
resolution\cite{bib:unal}  of
$\sigma(E) / E = 3.2\% /\sqrt{E} \oplus 9 \% / E \oplus 0.42 \%,$
where $E$ is in GeV.  The calorimeter is subdivided into 13248 cells of
transverse dimension 2\,cm $\times $ 2\,cm.


The reconstructed kaon mass in $K^\pm\rightarrow\pi^\pm\pi^0\,(K_{2\pi})$
decays has a typical  resolution of 3.25$\pm 0.05$\,MeV/c$^2$, without imposing
the $\pi^0$ mass constraint.

A scintillator hodoscope  is located between the spectrometer and the LKr
for triggering purposes. It
consists of two planes, segmented in horizontal and vertical strips
respectively,  with each plane arranged in four quadrants.  The time resolution
for the hodoscope system  is 200\,ps.

Downstream of the LKr calorimeter  is an
iron-scintillator sandwich hadron calorimeter\,(HAC), followed by muon
counters\,(MUC)  which consist of three planes of plastic scintillators
shielded by the HAC and a 80\,cm thick iron wall.  The first two planes
are made of
25\,cm wide horizontal  and 25\,cm wide vertical  scintillator strips, with a length of
2.7\,m.  The third plane consists of horizontal strips of width 44.6\,cm and
 is mainly used to measure the efficiency of the counters in the first two
planes.  The central strip in  each plane  is split with a gap of 21\,cm
to accommodate the beam pipe. The fiducial volume of the experiment  is
principally determined by the LKr calorimeter acceptance.

\subsection{Trigger and readout}

\label{sec:t_and_r}

The data coming from the detector  is   sampled every 25\,ns, and
eight samples are
recorded in  time windows of $200$\,ns.

The trigger selection  is
defined by at least one hit in each of the horizontal and
vertical planes of the hodoscope, within the same quadrant.
The typical rate  of this trigger at the reduced intensity  is
$1.3\times 10^{5}$ per burst, and is
downscaled by a factor of four in order to match   the
allowed limit of the data-acquisition system
of 50000 events per burst.
The efficiency of this trigger for events containing one track
 is calculated from events collected
by requiring at least three hit wires in at least three views of the upstream
chamber recorded within 200\,ns.
As shown in Table~\ref{tab:summary}, the trigger efficiency is found
to be high and independent of the decay mode and the type of kaon beamline.

\section{Event selection and reconstruction}
\label{sec:anal}

The signals for the $K_{e3}$, $K_{\mu3}$ and $K_{2\pi}$  decay modes require
one  track in the  drift chambers and at least two clusters
(photons) in the LKr that are consistent with a $\pi^0$ decay.
It is an important feature of this analysis that
their basic  event reconstruction and selection criteria are the same.  These   decay
modes are separated  from each other on the basis of
particle identification and additional kinematic
requirements, which also serves to reduce the background to a negligible level.

\subsection{Common selection criteria for $K_{e3}$, $K_{\mu3}$ and $K_{2\pi}$}
\begin{figure}[hbtp]
  \vspace{9pt}
  \centerline{
    \epsffile{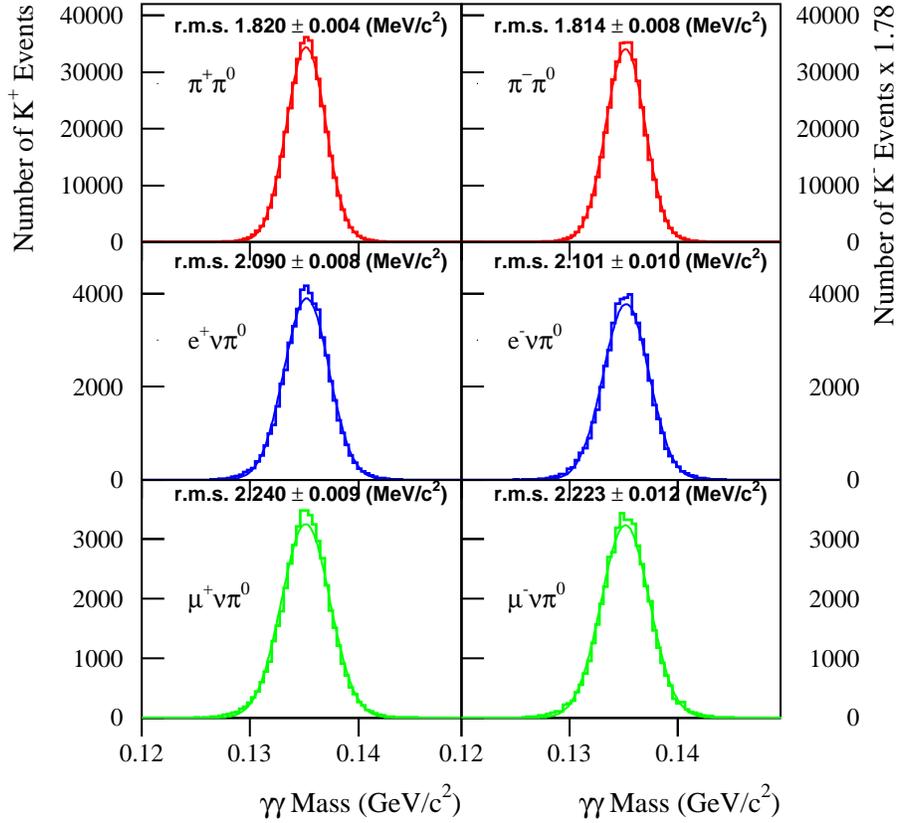}
}
\caption{ The three final states are characterized by
different average photon energies and acceptance. This fact is reflected
into the variation of the root mean square (r.m.s.) of the
$\gamma\gamma$ invariant mass distribution.
 The solid curves represent Gaussian fits  with
the particular r.m.s. value noted in each plot.
These variations however are small enough
and do not introduce systematics in the normalisation of the results.
 A clear $\pi^0$ signal is
shown in all channels after applying the full event selection.
The factor of 1.78 corresponds to the ratio of $K^+$ to $K^-$ flux in
the beamline.
\label{fig:pi0mass}}
\end{figure}

All events are required to have
at least one track that satisfies the following conditions:
\begin{itemize}
\item the track must be at least 15\,cm away
from the center of each drift chambers
and at least 15\,cm from the centerer of the LKr, 11\,cm from its outside
borders, and 2\,cm from any inefficient cell;
\item  to be within the allowed 37\,ns readout window defined by the trigger
hodoscopes;
\item to pass all track quality requirements.
\end{itemize}
The decay vertex, $V_{x,y,z}$, is reconstructed from the charged track and the
beam axis.
After correcting for  residual magnetic fields inside the decay pipe, it is
 required that:
\begin{itemize}
\item  the longitudinal position, $V_{z}$, must be 1316\,cm downstream of  the final
beam collimator,
and 2584\,cm upstream from the end of the decay pipe,  for
a total of  75\,m
of decay volume;
\item  the  transverse position, $V_{x,y}$, must be within
three  sigma (around 2.5\,cm) in $x$ and $y$
from the beam axis.  The actual cut
depends on the longitudinal vertex position due to the varying
uncertainty of the extrapolation to the vertex position.
\end{itemize}
All cluster  candidates in the $\pi^0$ reconstruction need to be clearly
identified as photons, that is:
\begin{itemize}
\item the energy of the cluster is greater than 3\,GeV and  less  than 65\,GeV;
\item time difference between clusters is smaller than 5\,ns;
\item the  minimum distance between clusters is 10\,cm in order
to minimize the effect of energy sharing on cluster reconstruction;
\item the  minimum distance between clusters and extrapolated tracks is greater
than 10\,(35)\,cm, so that no charged lepton\,(hadron) track is associated to the
clusters.
\end{itemize}
To make an inclusive
measurement of the radiation, a minimum of two reconstructed photons is
required. The threshold to define a reconstructed
photon, 3 GeV, leads to a high fraction of the events with only two
photons reconstructed. The fraction of events with at least one
 extra cluster in the calorimeter is
2.3\% in the case of $K_{e3}$, 0.3\% in $K_{\mu3}$ and 10\% in $K_{2\pi}$.
Fig.~\ref{fig:pi0mass} shows the invariant mass distribution for these two
photon events reconstructed using the $V_z$ component of the
charged vertex. The  non-Gaussian tails are of the same order in all three
channels. No requirement on the $\gamma\gamma$ invariant  mass is made.

For the small fraction of events with extra clusters,
the best  $\pi^0$ candidate is selected  by comparing  the $V_z$ component
of the charged vertex and   the longitudinal vertex, $z_\pi$, of the  $\pi^0$.
This neutral vertex is
reconstructed by assuming the $\pi^0$ mass and using the kinematical
information from the two photon candidates in the
$\pi^0\rightarrow \gamma\gamma$ decay.
$z_\pi$ is calculated from the  distance, $d_{\pi}$, of the $\pi^0$  vertex to
the LKr calorimeter  calculated from:
\begin{equation}
\label{eq:zk}
d_{\pi}={\sqrt{E_iE_j[(x_i-x_j)^2 +(y_i-y_j)^2]}}/{m_{\pi^0}} \, ,
\end{equation}
where $x_i$ and $y_i$ are the coordinates of the impact point of a the
$i^{th}$ $\gamma$  in the LKr, and $E_i$ is the energy of the corresponding
$\gamma$.  The longitudinal vertex position, $z_\pi$,
is given by \mbox{$z_\pi = z~(\mathrm{LKr~ position}) - d_{\pi}$}.
Although there is no explicit requirement on $V_z-z_\pi$, the combination with the
smallest
$V_z-z_\pi$ is taken.

\subsection{Selection criteria to distinguish between $K_{e3}$, $K_{\mu3}$ and
$K_{2\pi}$}
\begin{figure}[hbtp]
  \vspace{9pt}
  \centerline{
    \epsffile{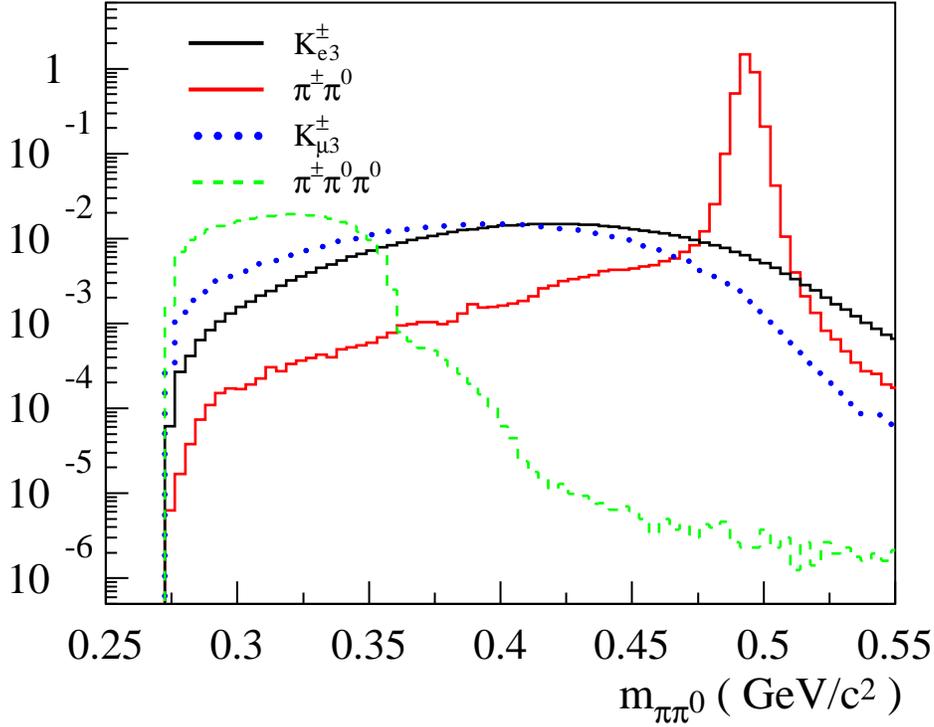}
}
\caption{Monte Carlo simulation for $K_{e3}$, $K_{\mu3}$, $K_{3\pi}$ and $K_{2\pi}$.
These are the distributions of the reconstructed invariant mass of the
two best photons and the track without applying particle identification,
and  under the assumption that the two photons come from a
$\pi^0$ decay and the track is a $\pi^\pm$.
\label{fig:mass}}
\end{figure}

The  invariant mass for
the $K_{2\pi}$  candidates  is required  to be within three  sigma of the reconstructed
$K^\pm$ mass, that is:
\begin{equation*}
\nonumber
0.4772~{\rm GeV/c^2}~<~m_{\pi^\pm\pi^0}~<~0.5102~{\rm GeV/c^2},
\end{equation*}
while  for $K_{\ell 3}$ candidates:
\begin{equation*}
\nonumber
m_{\pi^\pm\pi^0}~<~0.4772~{\rm GeV/c^2} ~~~{\rm or }~~~~ m_{\pi^\pm\pi^0}~>~0.5102~{\rm
GeV/c^2},
\end{equation*}
assuming $m_{\pi^\pm}$ instead of  $m_{\ell}$.
The Monte Carlo simulated  reconstructed invariant mass for the track
and the $\pi^0$, without applying particle identification and  under the assumption
that the track is a $\pi^\pm$, is shown in Fig.\,\ref{fig:mass} for the relevant decay
channels.

The allowed range in the transverse momentum
of the track, $P_{T}$,  with respect to the beam axis is required to be:
\begin{equation}
\nonumber
K_{e3}:~P_{T}~<~0.200~{\rm GeV/c},~~~~
K_{\mu 3}:~P_{T}~<~0.200~{\rm GeV/c},~~~~
K_{2\pi}:~P_{T}~<~0.215~{\rm GeV/c}.
\end{equation}

\begin{figure}[hbtp]
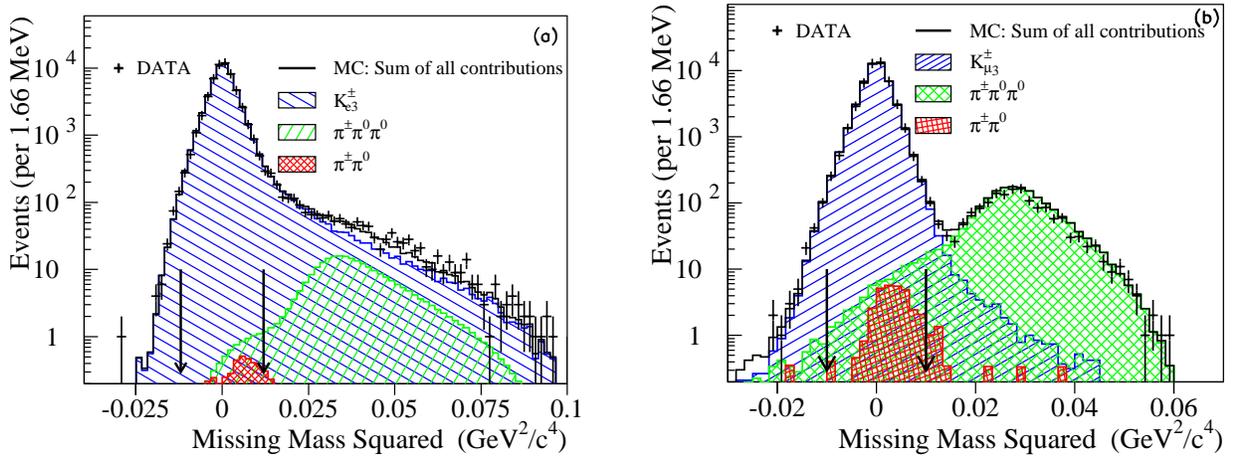

  \vspace{9pt}
  \centerline{\hbox{ \hspace{0.0in}
    \epsfxsize=3.0in
    \epsffile{plots/nu_ke3.epsi}
    \hspace{0.25in}
    \epsfxsize=3.0in
    \epsffile{plots/nu_kmu3.epsi}
}
}
\caption{Data and Monte Carlo comparison
for the reconstructed missing mass squared, $m_\nu^2$,
for (a) $K_{e3}$ and (b) $K_{\mu3}$ candidates. Only the information from the
charged lepton and the $\pi^0$ is used and extra
photons are ignored, if present in the event.
A good description is found after summing
the Monte Carlo prediction for
 the signal and the various background components.  Arrows indicate the allowed range
in
the
event selection.
The charged kaon energy is assumed to be
60 GeV.
\label{fig:neutrino_mass}}
\end{figure}

The missing mass squared, $m_\nu^2$,
is reconstructed under the assumption that the kaon energy is 60~GeV.
The allowed $m_\nu^2$ range for each channel is:
\begin{eqnarray*}
\nonumber
&K_{e3}:&~~-0.012~~{\rm GeV^2/c^4}<~m_\nu^2~<~0.012~{\rm GeV^2/c^4},\\
&K_{\mu 3}:&~~-0.010~~{\rm GeV^2/c^4}<~m_\nu^2~<~0.010~{\rm GeV^2/c^4},\\
&K_{2\pi}:&~~-0.0025~{\rm GeV^2/c^4}<~m_\nu^2~<~0.001~{\rm GeV^2/c^4},
\end{eqnarray*}
in order to distinguish between
events that are consistent with a three body or a two body decay,
see Fig.~\ref{fig:neutrino_mass}.

\subsection{Additional selection criteria for $K_{e3}$ candidates}
\begin{figure}[hbtp]
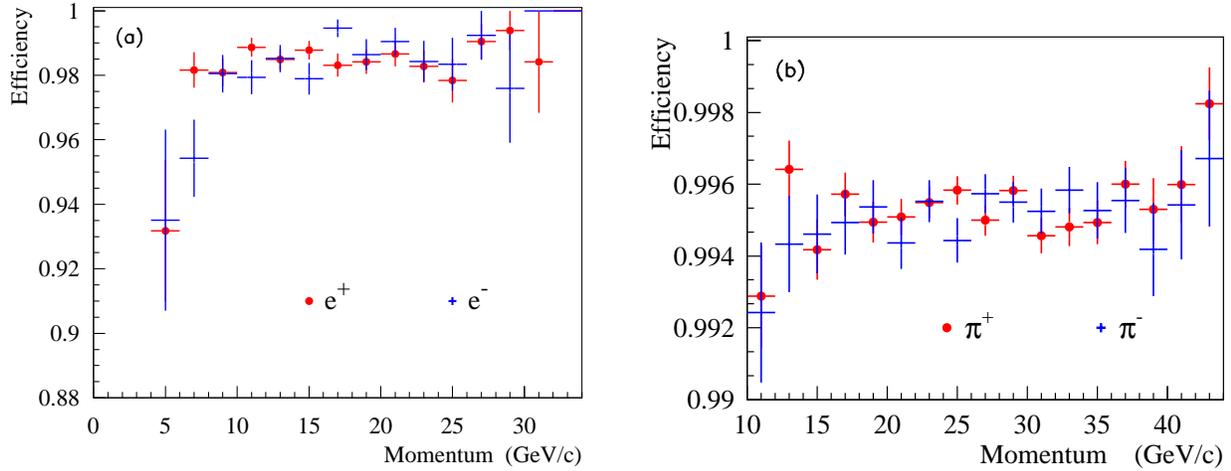

  \vspace{9pt}
  \centerline{\hbox{ \hspace{0.0in}
    \epsfxsize=3.0in
    \epsffile{plots/part_id_elect.epsi}
    \hspace{0.25in}
    \epsfxsize=3.0in
    \epsffile{plots/part_id_pions.epsi}
}
}
\caption{The $E/pc$ particle identification efficiency for (a)\,electrons and
(b)\,charged pions from clean subsamples of $K_{e3}$ and $K_{2\pi}$ decays
respectively.
These data samples have more stringent requirements on the
kaon mass, the missing mass and the
transverse momentum of the track.
\label{fig:eop}}
\end{figure}

The additional requirements for this channel are: (1)~the electron
identification is performed by combining the calorimeter energy measurement
($E$) with the spectrometer
momentum ($p$), and requiring $E/pc>0.95$.  The average  electron identification
 efficiency is
(98.59$\pm$0.09)\%, and the efficiency as a function of $p$
is shown in Fig.~\ref{fig:eop}. The error shown for the electron identification
is only statistical in nature, but the systematic uncertainty is expected to be small
since the result is  reproduced using
Dalitz decays; (2)~a minimum momentum of 5\,GeV/c is
required, in order to avoid low particle identification efficiency, and a
maximum momentum of 35\,GeV/c is also applied to
avoid regions of very low rate; (3)~the electron and the $\pi^0$
energies in the kaon center of mass are  required to be less than 0.22\,GeV and
0.27\,GeV
respectively,  to further reduce the background from
$K_{2\pi}$;
and (4)~$m_{\mu\pi^0}<0.425~{\rm GeV/c^2}$ is also required in order to
further  reduce the $\pi^\pm\pi^0$ background.

Using Monte Carlo simulations, the
backgrounds to this channel are  found to be well below the per mille level, see
Table~\ref{tab:background} and
 Fig.~\ref{fig:neutrino_mass}\,(a).
The main contribution comes  from decays with a $\pi^0$ and a $\pi^\pm$ that pass the
$E/pc$ requirement; for example, from
$K_{2\pi}$ that has a branching fraction that is about  four times larger, and
$K_{\pi^\pm\pi^0\pi^0}$ where
only the information from two of the photons are used.

\subsection{Additional selection criteria for $K_{\mu3}$ candidates}
\begin{figure}[hbtp]
  \vspace{9pt}
  \centerline{
\epsfxsize=3.0in
    \epsffile{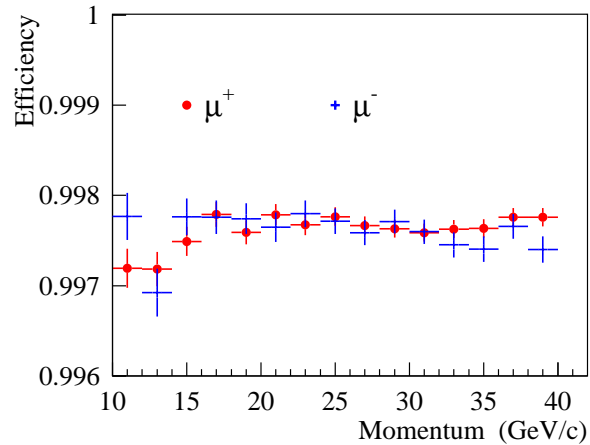}
}
\caption{
 Particle identification efficiency for muons measured using a sample
obtained from $K^\pm\rightarrow \mu^\pm\nu$ decays.
\label{fig:muon}}
\end{figure}

The additional requirements for this channel are: (1)~in order to identify
a muon a charged track, reconstructed in the spectrometer, is
extrapolated to the MUC planes and associated with the MUC hits after testing
the spatial separation
and time difference between this extrapolated track and the MUC hits.
Multiple scattering is taken into account before applying the spatial cut,
and light propagation along the MUC strips is taken into account
before applying the time-difference cut. A track is classified as a muon if the
following conditions are  satisfied:
\begin{itemize}
\item each track has hits in  planes one and two of the MUC;
\item for each track, the  time difference  measured by the chambers or
the hodoscope, and by the muon detector does not differ by
more than 4.5\,ns;
\end{itemize}
Using these requirements, the efficiency for muon identification is
measured from a kinematically selected sample of $K\rightarrow \mu\nu$
and found to be larger than  0.995 as shown in Fig.\,\ref{fig:muon}.
The average muon identification efficiency is (99.759$\pm$0.003)\%.
The error shown for the muon identification is only statistical in nature,
but the systematic uncertainty is expected to be small since the result is
reproduced using data taken  with a dedicated muon beam;
(2)~The individual track momenta is required to be
above 10~GeV/c and below 40~GeV/c;
(3)~the energy of the  muon  in the  kaon center of mass is below 0.23\,GeV;
(4)~the $\pi^0$ energy in the kaon center of mass  is below 0.23\,GeV or
$m_{\mu\pi^0}~<~0.38~{\rm GeV/c^2}$; and
(5)~the $\pi^0$ energy is smaller than  40~GeV  to reduce the background from
$K_{2\pi}$.

Backgrounds to this channel are  found to be at the two per mille level, see
Table~\ref{tab:background}
and Fig.~\ref{fig:neutrino_mass}\,(b).
The main contribution comes from decays with a $\pi^0$ together with a $\pi^\pm$ that
decays in flight.

\subsection{Additional selection criteria for $K_{2\pi}$ candidates}
The additional requirements  for this channel are: (1)~muons are  not rejected
to avoid losing events where the pion decays
(this corresponds to 1.6\% of the $K_{2\pi}$ events), but electrons are
rejected by requiring $E/pc<0.95$. The average pion identification
efficiency is (99.524$\pm$0.009)\%. The fraction
of surviving pions after this requirement is shown in Fig.~\ref{fig:eop}(b).
 The error shown for the pion identification is only statistical in nature,
but the systematic uncertainty is expected to be small since the result  at low
momenta is reproduced using an independent sample of $\pi^\pm\pi^\pm\pi^\mp$
decays.
(2)~The individual track momenta are  required to be above 10~GeV/c and below
50~GeV/c.

Backgrounds to this channel are  found to be at the few
 per mille level,
see Table~\ref{tab:background}.  The main contribution is from  $K_{\mu3}$
because we do not reject events with muons.

\subsection{Final data samples}
Around 56K\,(31K), 49K\,(28K) and 462K\,(256K) events  are found for
$K_{e3}^+\,(K_{e3}^-)$, $K_{\mu3}^+$ $(K_{\mu3}^-)$ and
$K_{2\pi}^+\,(K_{2\pi}^-)$ candidates respectively,
after all the event selection requirements are applied
(see
Table~\ref{tab:summary}).

\section{Acceptance calculations}

\label{sec:extraction}

To obtain the acceptance and an estimate of  the background
fractions, a GEANT\cite{bib:geant} based Monte Carlo simulation  is
modeled for each beam  separately.  The beam optics for the $K^+$ and
the $K^-$ beamlines are  described using TURTLE~\cite{bib:turtle}, and fine
tuned
using fully reconstructed $\pi^\pm\pi^\pm\pi^\mp$ and $\pi^\pm\pi^0$ events.

The main inputs needed to describe the decay amplitude for
semileptonic decays are: (1) the radiative corrections, and (2) the
parameters,  $\lambda'_+$, $\lambda''_+$ and $\lambda_0$, of  the  model presented in
 Eq.\,(\ref{eq:q2})  and used to
describe the dependence of the form factors, $f_+$ and $f_0$,  on the momentum
transferred to the leptons, $t$; see Section~\ref{sec:phenom}.

In order to reduce the uncertainties due to radiative effects,  after
identifying   the track and the $\pi^0$, events with
extra photons are  kept. Nevertheless,
it is important that the Monte Carlo includes a well defined fraction of radiative
decays (like $K_{e3\gamma}$) in order to
obtain an acceptance correction that is  accurate at the percent level.
For example, studies show that ignoring the radiative events in   the
$K_{e3}$ generation causes the acceptance to be overestimated by 1.6\%,
even after correcting  for virtual
effects on the three-body Dalitz plane region using the prescription given
in~\cite{bib:gin}.

\begin{figure}[hbtp]
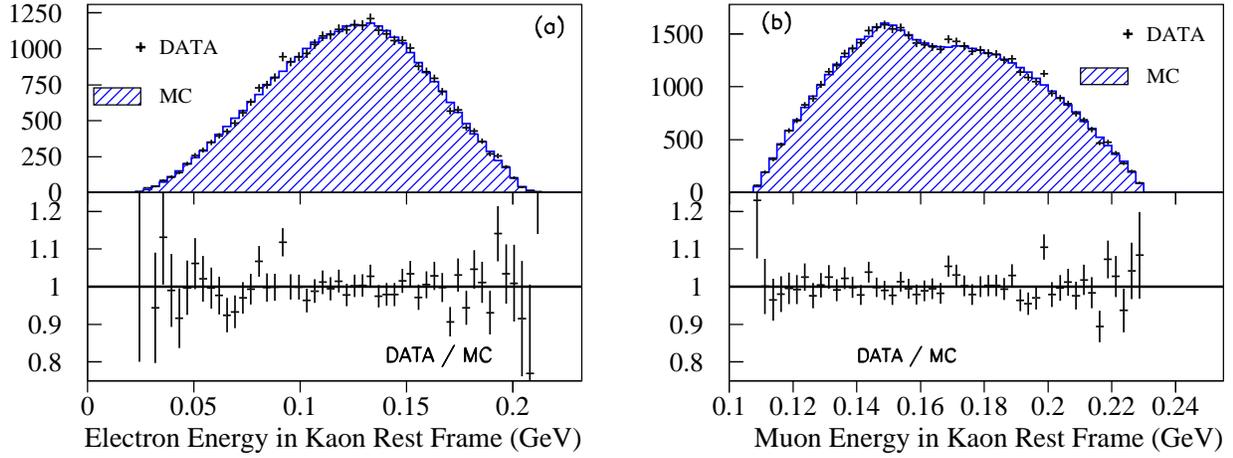

  \vspace{9pt}
  \centerline{\hbox{ \hspace{0.0in}
    \epsfxsize=3.0in
    \epsffile{plots/com_trk_ke3m.epsi}
    \hspace{0.25in}
    \epsfxsize=3.0in
    \epsffile{plots/com_trk_kmu3.epsi}
}
}
\caption{
     Comparison of the (a)\,electron and (b)\,muon energy in the kaon rest
frame for $K_{e3}$ and $K_{\mu3}$  events.
     Ratio of the Data over Monte Carlo is given in the lower plots.
\label{fig:mc_ke3}}
\end{figure}

\begin{figure}[hbtp]
  \vspace{9pt}
  \centerline{
    \epsffile{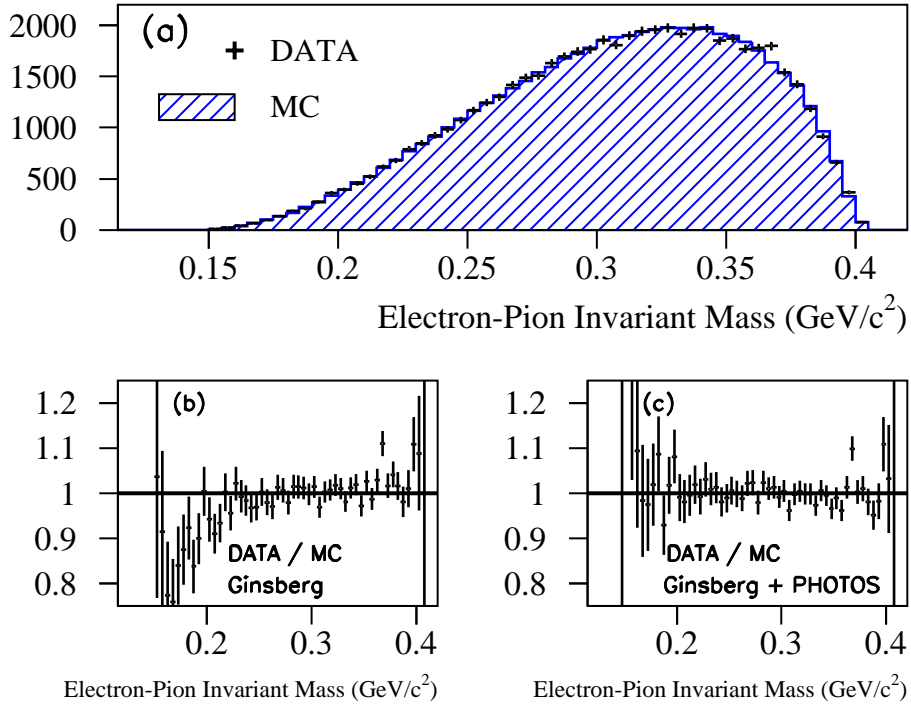}
}
\caption{
     (a) Comparison of the electron and the $\pi^0$ invariant mass distribution
for $K_{e3}$ events.  Ratio of the Data over Monte Carlo  given in the lower plots for
Monte Carlo generated with corrections to the Dalitz distributions for virtual
effects\cite{bib:gin}
and real photons due to  bremsstrahlung\cite{Photos}
(c) is in good
agreement, while
(b) only correcting the Dalitz distributions for virtual effects\cite{bib:gin}
and  ignoring the $K_{e3\gamma}$ contribution fails to  described the
low mass region.
\label{fig:ke3g}}
\end{figure}

The PHOTOS
package \cite{Photos} is used to simulate bremsstrahlung, and
calculations from \cite{bib:gin}  are added to include virtual photons and
electrons.  Figures~\ref{fig:mc_ke3}(a) shows a comparison
between data and Monte Carlo events for the electron energy in
the center of mass of the kaon.
Good agreement is found, even though this distribution is  sensitive
to radiative effects.
All other kinematic distributions show that the data are well described by the
Monte Carlo after including effects introduced by $K_{e3\gamma}$ events. For example,
the invariant mass of the electron and $\pi^0$ pair is sensitive to
$K_{e3\gamma}$, see Fig.~\ref{fig:ke3g}.
 The proper description of the low mass region
proved to be sensitive to the inclusion of the  $K_{e3\gamma}$.
The effect of radiative corrections to the  $K_{e3}$ acceptance
is confirmed using  the
program described in reference~\cite{bib:kloe} that includes virtual corrections
and the  corresponding radiative decay simultaneously.
The  $K^+_{e3}(K^-_{e3})$  acceptance is found to be
$0.0709\pm 0.0001\,(0.0706\pm 0.0001)$.

For $K_{\mu3}$, the  prescription given in \cite{bib:gin} is sufficient
due to the
smaller  bremsstrahlung contribution, and adding PHOTOS had a negligible impact
on the acceptance.
Fig.~\ref{fig:mc_ke3}(b) shows a comparison between data and Monte Carlo events
for the muon   energy in the center of mass of the kaon.
The  $K^+_{\mu3}(K^-_{\mu3})$ acceptance is found to be
$0.0930\pm 0.0001\,(0.0927\pm 0.0001)$.

For a similar reason, the change in acceptance in $K_{2\pi}$ is found
to be smaller than one~per~mille if the radiative events are included in the
generator.  The  $K^+_{2\pi}(K^-_{2\pi})$  acceptance is found to be
$0.1424\pm 0.0001\,(0.1419\pm 0.0001)$.

The response of the LKr detector, even though simulated, is  not
used  for particle identification during the acceptance calculations.
The particle identification efficiency
is studied using data and corrections are made separately, see
Fig.\,\ref{fig:eop} and \ref{fig:muon}.
The acceptances after including the corrections due to particle
identification for all three decay modes are summarized
in Table~\ref{tab:summary}, and the assigned uncertainties are statistical in nature.

The acceptance for $K^+$ is systematically higher than for $K^-$ due to
differences introduced by the polarity selected for the spectrometer
magnet.

\subsection{Form factor description in the Monte Carlo generation}
\label{sec:extraction_ff}
The Dalitz distributions for the semileptonic decays are generated with:
$\lambda'_+=0.02485\pm 0.00163\pm0.00034$,
$\lambda''_+=0.00192\pm 0.00062\pm 0.00071$ (both from $K_{e3}$
decays)\,\cite{bib:PDG},
and $\lambda_{0}=0.0196\pm0.0012$\,(assuming  $\mu-e$ universality)\,\cite{bib:PDG}.
The change in acceptance when the current values of $\lambda'_+$,
$\lambda''_+$ and $\lambda_0$ are changed by one sigma is
about 0.08\%\,(0.07\%) for  $K_{e3}$\,($K_{\mu3}$).
The correlation between $\lambda_{+}^{'}$ and $\lambda_{+}^{''}$ is assumed to
be -0.95\,\cite{bib:PDG}, while the measurements for the vector and the scalar
form factor parameters are assumed to be uncorrelated.
Table~\ref{tab:slope_result_ff_par} shows the
the  resulting slope in
the  \rkekp,~ \rkmukp ~and~ \rkmuke ~measurements after changing  form factor
parameter by $\pm$ one sigma, while maintaining the other parameters
constant.
The contribution
from this change in acceptance to the total systematic error is included in
Table~\ref{tab:ke3_ku3_ku3ke3_sys}.

The changes in acceptance due to alternative models describing the $t$
dependence of the form factors are given below:
\begin{itemize}
\item the linear approximation:
\begin{equation}
\label{eq:lin}
f_{+,0} (t) = f_+ (0)\left(1+\lambda_{+,0} {\frac{\textstyle t}{\textstyle
m_{\pi^\pm}^2}} \right) \; ,
\end{equation}
where $\lambda_{+}=0.0296\pm 0.0008$  and $\lambda_{0}=0.0196\pm0.0012$
(assuming  $\mu-e$ universality)\,\cite{bib:PDG}.  In this case, the ratio
between the acceptance obtained using this approximation compared to the
acceptance using the quadratic approximation (Eq.~\ref{eq:q2}) is found to be
1.0006   and 0.9998 for $K_{e3}$ and $K_{\mu3}$, respectively.
\item the pole approximation:
\begin{equation}
f_{+,0} (t) = f_+ (0)\left(\frac{m_{V,S}^2}{m_{V,S}^2-t}\right) \; ,
\end{equation}
where
$m_V=0.877\pm 0.005\,{\rm GeV/c^2}$ for  $K_{e3}^{0}$ and $K_{\mu3}^{0}$
(assuming $\mu-e$ universality)\,\cite{bib:PDG}
 and $m_S=1.187\pm 0.050\,{\rm GeV/c^2}$
for $K_{\mu3}^{0}$ (assuming $\mu-e$ universality)\,\cite{bib:PDG}.
These pole masses are obtained from neutral kaons.
In this case, the ratio between the acceptance obtained using this
approximation compared to the acceptance using the quadratic approximation
(Eq.~\ref{eq:q2}) is found to be
0.9996   and 0.9984 for $K_{e3}$ and $K_{\mu3}$, respectively.
\end{itemize}
The full difference in acceptance between the pole approximation and the quadratic
approximation
is assigned as an uncertainty due to the choice of form factor model,
see Table~\ref{tab:ke3_ku3_ku3ke3_sys}.

\section{Result}
\label{sec:results}

Table~\ref{tab:summary} lists all
quantities  needed to evaluate \mbox{${\cal R}_{K_i/ K_j}$}:
\begin{equation}
{\cal R}_{K_i/ K_j}=\frac{
Acc_{K_{j}}\times\epsilon_{track_{ID_j}}\times Trig_{K_{j}}\times N_{K_{i}} \times
(1+\Delta_{K_{j}})}
{
Acc_{K_{i}}\times\epsilon_{track_{ID_i}}\times Trig_{K_{i}}\times N_{K_{j}} \times
(1+\Delta_{K_{i}})}\, ,
\end{equation}
where $i,j= \ell 3,2\pi$.
The correction to $K_{e3}$ due to particle identification efficiency
($\epsilon_{track_{ID}}$)
amounts to a few percent, while the background correction  $ (1+\Delta_{K_{i}})$ is
negligible.
$K_{\mu3}$ and  $K_{2\pi}$ require a particle identification
correction that is below the percent level, and a correction
for background correction  that is only a few per mille.

The  results for $K^+$ and $K^-$ combined are:
\begin{eqnarray}
   \rkekp  &=& 0.2470 \pm 0.0009 (stat)\pm 0.0004 (syst),\\
   \rkmukp &=& 0.1637 \pm 0.0006 (stat)\pm 0.0003 (syst),\\
\label{eq:kmuke}
   \rkmuke &=& 0.663  \pm 0.003  (stat)\pm 0.001   (syst).
\end{eqnarray}
The individual $K^+$ and $K^-$ results and their systematic uncertainties are
listed in Table~\ref{tab:ke3_ku3_ku3ke3_sys}.  The sources of systematic uncertainties
are due to the corrections listed in Table~\ref{tab:summary} and the
treatment of the form factors as discussed in Section~\ref{sec:extraction_ff}.

The ratios are found to be insensitive to the photon reconstruction and track-finding.
Analysis of these ratios as a function of their basic distributions show stability.

\begin{figure}[hbtp]
  \vspace{9pt}
  \centerline{
\epsfxsize=3.0in
    \epsffile{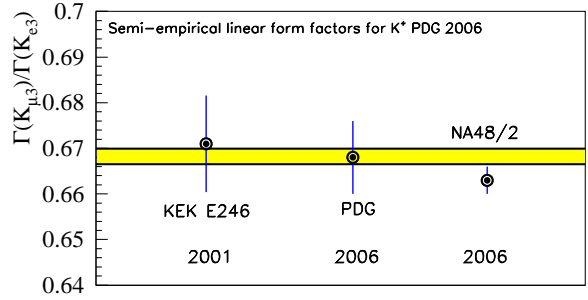}
}
\caption{
 $\rkmuke$ results compared to  KEK-246 results~\cite{bib:kek246},
the corresponding PDG value of
2006~\cite{bib:PDG}  and
to the predictions
assuming $\mu-e$ universality, Eq.~(\ref{eq:ratio}),  with the $\lambda_+$ and
$\lambda_0$ values
given for $K^{\pm}$ in the PDG of
2006~\cite{bib:PDG}.
\label{fig:results2}}
\end{figure}

The final results are shown in Fig.~\ref{fig:results1}
for  \rkekp\ and \rkmukp, and in Fig.~\ref{fig:results2} for \rkmuke.
These can be compared to the
 current PDG values of
$ \rkekp= 0.238\pm  0.004$\,\cite{bib:PDG},  \rkmukp~$=
0.159 \pm 0.003$\,\cite{bib:ratio_kmu3_pi} and
$\rkmuke=0.668\pm0.008$\,\cite{bib:PDG}.
Taking the current PDG value for the $K_{2\pi}$  branching fraction,
$0.2092\pm 0.0012$\,\cite{bib:PDG},
the branching fractions for the semileptonic decays are found  to be:
\begin{eqnarray}
\label{eq:br_results1}
Br(K_{e3})  &=&0.05168\pm 0.00019(stat)\pm 0.00008(syst)\pm 0.00030(norm),\\
\label{eq:br_results2}
Br(K_{\mu3})&=&0.03425\pm 0.00013(stat)\pm 0.00006(syst)\pm 0.00020(norm).
\end{eqnarray}
The uncertainty is dominated by the existing data for the $K_{2\pi}$ branching
fraction.  Recall the corresponding PDG values\,\cite{bib:PDG} are
$Br(K_{e3}) = 0.0498\pm0.0007$ and $Br(K_{\mu3}) = 0.0332\pm0.0006$.
Higher branching fractions are found for both $K_{e3}$ and $K_{\mu3}$,
confirming the $K_{e3}$ results reported by the BNL-E865
collaboration\,\cite{bib:bnl}.

\section{Discussion}
\label{sec:dis}

\subsection{$V_{us}$ matrix element}
As discussed in Section~\ref{sec:phenom}, the measured partial widths
for semileptonic decays can be combined with theoretical corrections and
other experimental measurements to calculate $|V_{us}|$.

Using the newly measured  branching fractions given in
Eqs.~(\ref{eq:br_results1})-(\ref{eq:br_results2}), the $K^\pm$ lifetime
$\tau^{PDG}_{K^+}=(1.2385\pm 0.0024)\times 10^{-8}\,s$\,\cite{bib:PDG},
$G_F= (1.16637\pm0.00001)\times 10^{-5}$\,GeV$^{-2}$\,\cite{bib:marciano},
$m_K=0.493677\pm 0.000016$\,GeV/c$^2$\,\cite{bib:PDG},
$S_{EW}=1.0230\pm0.0003$\,\cite{bib:sirlin},
 and
the full  phase space integrals and long-distance corrections as  given in
Table~\ref{tab:vus_input},
the $|V_{us}|$  matrix element times the vector form factor $f_{+}(0)$ is
found to be (see Eq.~(\ref{eq:vus})):
\begin{eqnarray}
\label{eq:vf1}
|V_{us}|f_{+}(0) &=& 0.2193  \pm 0.0012 \,,~~~~~~~~~[K_{e3}]\\\nonumber
                 &=& 0.21928 \pm 0.00039(stat) \pm  0.00016(syst) \pm 0.00062(norm) \pm
0.00095(ext),\\
\label{eq:vf2}
                 &=& 0.2177  \pm 0.0013 \,,~~~~~~~~~[K_{\mu3}] \\\nonumber
                 &=& 0.21774 \pm 0.00041(stat) \pm 0.00019(syst) \pm 0.00064(norm) \pm
0.00103(ext),
\end{eqnarray}
from $K_{e3}$ and $K_{\mu3}$, respectively.  53\%\,(60\%) of the external ($ext$) error
is due to the
sum of the
long-distance corrections, $SU(2)$ and $EM$ corrections, taken
from~\cite{bib:Cirigliano,bib:gin,bib:wgr} for $K_{e3}$ and $K_{\mu3}$, respectively.
The phase space integral, Eq.~(\ref{eq:int}), is  evaluated using the quadratic
approximation,
Eq.~(\ref{eq:q2}), and gives the next largest contribution to the external error.
The last significant uncertainty comes from the error in the $K^\pm$ lifetime.
Combining these $|V_{us}|f_{+}(0)$ values by     assuming
$\mu-e$ universality,  we obtain:
\begin{eqnarray}
|V_{us}|f_{+}(0) &=& 0.2188  \pm 0.0012 \,,
\label{eq:vusresult}
\\\nonumber
   |\vus| &=& 0.2277 \pm 0.0013{\,(other)} \pm 0.0019 {\,(theo)},
\end{eqnarray}
where $``theo"$ refers to the theoretical uncertainty due to $f_+(0)$, and
$``other"$ refers to all the uncertainties already included in
Eqs.\,(\ref{eq:vf1})-(\ref{eq:vf2}) and their correlation.
To extract $|V_{us}|$, it is  assumed that the value of $f_+(0)$
is $0.961 \pm 0.008$\,\cite{bib:Leu_Roos}  as calculated for
neutral kaons. There is no need to make a distinction between charged and
neutral kaons in $f_+(0)$ because the $SU(2)$ correction is  applied directly,
see Eq.~(\ref{eq:vus})-(\ref{eq:int}) and Table~\ref{tab:vus_input}.

\begin{figure}
  \vspace{9pt}
  \centerline{
\epsfxsize=3.0in
    \epsffile{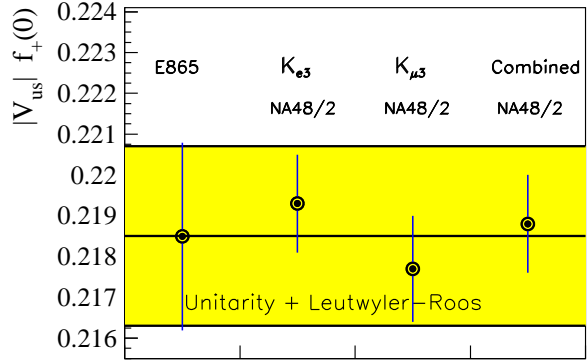}
}
\caption{
Comparison of the NA48 measurement of $|V_{us}|f_+(0)$ from $K_{e3}$ and
$K_{\mu3}$ data, and the $K_{e3}$  BNL-E865 result\,\cite{bib:bnl}.
The theoretical prediction shown is obtained assuming  unitarity  of
the CKM matrix $V_{ud}$,
using the PDG values for $V_{ud}$ and $V_{ub}$ as input, and using
the choice of $f_+(0)$ as described in the text.
\label{fig:results3}}
\end{figure}

These $|\vus|f_+(0)$  and $|\vus|$ values are
to be compared to predictions obtained by imposing the
unitary condition on the CKM matrix, and taking the latest values of
$|V_{ud}|=0.9738\pm0.0003$~\cite{bib:marciano2} and
$|V_{ub}|=(3.6\pm 0.7)\times 10^{-3}$\cite{bib:PDG}:
\begin{eqnarray}
   |\vus|_{unitary} f_+(0)&=& 0.2185  \pm 0.0022,\\
   |\vus|_{unitary}       &=& 0.2274  \pm 0.0013.
\end{eqnarray}
As shown in Fig.~\ref{fig:results3}, these theoretical  predictions are in good
agreement
with the results presented in Eq.~(\ref{eq:vusresult}).
The result is in good agreement with the unitarity of the CKM mass-mixing matrix.

\subsection{$\Gamma(K_{\mu3})/\Gamma(K_{e3})$}

The consistency between the width ratio for the semileptonic decays
already presented in Eq.\,(\ref{eq:kmuke}) and the world average for the linear
form factors  parameters,
$\lambda_+$ and $\lambda_0$\,\cite{bib:PDG},  can be tested.   The band in  Fig.~\ref{fig:results2} corresponds to
the predictions for $\rkmuke$ assuming $\mu-e$ universality, Eq.~(\ref{eq:ratio}), with the $\lambda_+$ and
$\lambda_0$ values given for $K^{\pm}$ in the PDG of 2006~\cite{bib:PDG}.
Using Eq.\,(\ref{eq:ratio}),  $\rkmuke$ is predicted   to be
$0.6682\pm 0.0017$.  The result for  $\rkmuke$ suggests a lower value for
$\lambda_{0}$ than the current world average for $K^{\pm}$\,\cite{bib:PDG}, 
as found in recent measurements from $K_L$\,\cite{bib:PDG} decays.

$\mu-e$ universality can be tested using the \rkmuke\ ratio and
Eq.\,(\ref{eq:no_lepton1}).  The result shown in Eq.\,(\ref{eq:kmuke})
implies that $g_{\mu}f_{+}^{\mu}(0)/g_{e}f_{+}^{e}(0)$ is  $0.99\pm0.01$,
which is consistent with unity within the experimental error.

\section*{Conclusion}

The measured ratios of the decay rates for the charged kaon semileptonic decays
are found to be:
\begin{eqnarray}
   \rkekp  &=& 0.2470 \pm 0.0009 (stat)\pm 0.0004 (syst),\\
   \rkmukp &=& 0.1637 \pm 0.0006 (stat)\pm 0.0003 (syst),\\
   \rkmuke &=& 0.663  \pm 0.003  (stat)\pm 0.001   (syst).
\end{eqnarray}

Using the current experimental knowledge of the $K_{2\pi}$ branching ratio,
this
leads to branching fractions of $Br(K_{e3})= 0.05168\pm 0.00036$ and
$Br(K_{\mu3})=0.03425\pm 0.00024$.  This exceeds the PDG value\,\cite{bib:PDG}
in both cases.

Combining these results, we find $|V_{us}| f_+(0) = 0.2188 \pm 0.0012$, in good
agreement with the
CKM unitary prediction.

$\rkmuke$ is in reasonable agreement with the semi-empirical
predictions and the error is improved compared to the current world average.

\section*{Acknowledgments}
It is a pleasure to thank the technical staff of the participating laboratories,
universities and affiliated computing centers for their efforts in the
construction of the NA48 apparatus, in the operation of the experiment, and in
the processing of the data.  We would also like to thank Gino Isidori and
    Vincenzo Cirigliano
for important discussions and  Claudio Gatti for his important contribution to the
calculation of radiative corrections.

\begin{table}
\begin{center}
\begin{tabular}{|c|c|c|c|c|}
\hline
\hline
Decay type &  Raw number   &  Acceptance $\times$   &   Backgrounds/Signal & Trigger \\
           &  of events    &  particle ID           &                      & Efficiency
\\
           & $(N_i)$       & $(Acc_i\times \epsilon_{track_{ID}})$&  $(\Delta_{i})$
&$(Trig_{i})$ \\
\hline
\hline
$K^+_{e3}$   &  56,196 & $ 0.0698\pm 0.0001 $ & $ (0.0200\pm 0.0008)\% $ & $ 0.9990 \pm
0.0005$\\
$K^-_{e3}$   &  30,898 & $ 0.0694\pm 0.0001 $ & $ (0.0209\pm 0.0010)\% $ & $ 0.9982 \pm
0.0008$\\
\hline
$K^+_{\mu3}$ &  49,364 & $ 0.0927\pm 0.0001 $ & $ (0.2215\pm 0.0079)\% $ & $ 0.9986 \pm
0.0006$\\
$K^-_{\mu3}$ &  27,525 & $ 0.0925\pm 0.0001 $ & $ (0.2175\pm 0.0077)\% $ & $ 0.9988 \pm
0.0007$\\
\hline
$K^+_{2\pi}$ & 461,837 & $ 0.1418\pm 0.0001 $ & $ (0.2893\pm 0.0058)\% $ & $ 0.9987 \pm
0.0002$\\
$K^-_{2\pi}$ & 256,619 & $ 0.1412\pm 0.0001 $ & $ (0.2896\pm 0.0058)\% $ & $ 0.9990 \pm
0.0002$\\
\hline
\hline
\end{tabular}
\hspace*{0.5cm}
\caption{Summary of information used to extract the branching ratio,
where $track=e^\pm, \mu^\pm, \pi^\pm$ for $i=K^\pm_{e3}, K^\pm_{\mu 3},
K^\pm_{2\pi}$.
\label{tab:summary}}
\end{center}
\end{table}

\begin{table}
\begin{center}
\begin{tabular}{|c|c|c|}
\hline
\hline
 Contributing channel &  $K^+$ & $K^-$  \\
\hline
\hline
\multicolumn{3}{|c|}{            $K_{e3}$ }   \\
\hline
$K_{\pi^\pm\pi^0\pi^0}$    & $(0.0130\pm0.0007)\% $   & $(0.0139\pm0.0009 )\%$ \\
$K_{2\pi}$                 & $(0.0070\pm0.0003)\% $   & $(0.0071\pm0.0004 )\%$ \\
\hline
\multicolumn{3}{|c|}{            $K_{\mu3}$ } \\
\hline
$K_{\pi^\pm\pi^0\pi^0} $   & $(0.1598 \pm 0.0045)\% $ & $(0.1599\pm 0.0046)\%$ \\
$K_{2\pi}              $   & $(0.0617 \pm 0.0064)\% $ & $(0.0576\pm 0.0061)\%$ \\
\hline
\multicolumn{3}{|c|}{            $K_{2\pi}$ } \\
\hline
$K_{\mu3}$                 & $ (0.2848\pm0.0058)\% $  & $(0.2846\pm0.0058)\%$ \\
$K_{e3}$                   & $ (0.0045\pm0.0006)\% $  & $(0.0050\pm0.0008)\%$ \\
\hline
\hline
\end{tabular}
\hspace*{0.5cm}
\caption{Percentage  of background
 expected from Monte Carlo simulation for  $K_{e3}$,  $K_{\mu3}$ and  $K_{2\pi}$ from
the main
contributors to their total background.
\label{tab:background}}
\end{center}
\end{table}

\begin{table}
\begin{center}
\begin{tabular}{|c|c|c|c|c|}
\hline
\hline
Parameter Changed  &  $\D{\lambda_{i}}{\rkekp} $ &
 $\D{\lambda_{i}}{\rkmukp}$   &  $ \D{\lambda_{i}}{\rkmuke}$  \\
\hline
\hline
$ \lambda_{+}^{'} $ & -0.160& -0.076 &~0.113 \\
$ \lambda_{+}^{''}$ & -0.319& -0.204 &~0.021 \\
$ \lambda_{0}     $ & ----- & -0.048 &-0.192 \\
\hline
\hline
\end{tabular}
\caption{{Dependence of the \rkekp,~ \rkmukp ~and~ \rkmuke ~measurements, when changing
the form factor
parameter, $\lambda_{i}$, by $\pm$ one sigma, and maintaining the other form factor
parameters constant.}}
\label{tab:slope_result_ff_par}
\end{center}
\end{table}

\begin{table}
\begin{center}
\begin{tabular}{|l|c|c|c|c|c|c|}
\hline
\hline
      &\multicolumn{2}{c|}{\rkekp} & \multicolumn{2}{c|}{\rkmukp} &
\multicolumn{2}{c|}{\rkmuke} \\
                      &  $K^+$  &  $K^-$ &  $K^+$   &   $K^-$ &  $K^+$  &  $K^-$ \\
\hline
\hline
Central value                  & 0.2476  & 0.2460  & 0.1636  & 0.1639  & 0.6605 &
0.6661 \\
\hline
Statistical error              & 0.0011  & 0.0015  & 0.0008  & 0.0010  & 0.0040 &
0.0055 \\
\hline
Total systematic error         & 0.0005  & 0.0006  & 0.0004  & 0.0004  & 0.0016 &
0.0018\\
\hline
Accept. $\times$ Part-ID num.  & 0.00041 & 0.00047 & 0.00017 & 0.00017 & 0.00067 &
0.00068 \\
Accept. $\times$ Part-ID denom.& 0.00021 & 0.00021 & 0.00014 & 0.00014 & 0.00106 &
0.00126 \\
Trigger efficiency in num.     & 0.00013 & 0.00020 & 0.00010 & 0.00011 & 0.00039 &
0.00046 \\
Trigger efficiency in denom.   & 0.00005 & 0.00005 & 0.00003 & 0.00003 & 0.00033 &
0.00053 \\
Background subtraction         & 0.00001 & 0.00001 & 0.00002 & 0.00002 & 0.00005 &
0.00005 \\
Uncertainty in form factor &\multicolumn{2}{c|}{0.00010} & \multicolumn{2}{c|}{0.00010}
& \multicolumn{2}{c|}{0.00029}\\
Uncertainty in f. f. model &\multicolumn{2}{c|}{0.00009} & \multicolumn{2}{c|}{0.00026}
& \multicolumn{2}{c|}{0.00086}\\
\hline
\hline
\end{tabular}
\hspace*{0.5cm}
\caption{Summary of the statistical and systematic uncertainties for the \rkekp,
\rkmukp~and \rkmuke~measurements.
\label{tab:ke3_ku3_ku3ke3_sys}}
\end{center}
\end{table}

\begin{table}
\begin{center}
\begin{tabular}{|l|c|c|c|c|c|}
\hline
\hline
Decay &Branching &  Phase Space& \multicolumn{2}{c|}{ Radiative } & $|V_{us}|f_+(0)$ \\
Channel &Fraction   &  Integral & \multicolumn{2}{c|}{
Correction\cite{bib:Cirigliano,bib:gin,bib:wgr}} &  \\
&$Br$ & $I_K^\ell$ & $\delta_{SU(2)}^\ell(\%)$ & $\delta^\ell_{EM}(\%) $ &  \\
\hline
\hline
$K_{e3}$&$0.0517\pm0.0004$   &$0.1591\pm 0.0012$ &$2.31\pm 0.22$&$0.03\pm
0.10$&$0.2204\pm 0.0012$\\
\hline
$K_{\mu3}$& $0.0343\pm0.0002$&$0.1066\pm 0.0008$&$2.31\pm 0.22$&$0.20\pm
0.20$&$0.2177\pm 0.0013$\\
\hline
\hline
\end{tabular}
\hspace*{0.5cm}
\caption{Inputs to Eq.~(\ref{eq:vus}) and results for $|V_{us}|f_+(0)$. By assuming
unitarity,
the prediction of  $|V_{us}|f_+(0)$ for charged kaons  is $0.2185  \pm 0.0022$.
\label{tab:vus_input}}
\end{center}
\end{table}

\begin{figure}[hbtp]
  \vspace{9pt}
  \centerline{\hbox{ \hspace{0.0in}
    \epsfxsize=2.75in
    \epsffile{plots/final_r_ke3_pipi0.epsi}
    \hspace{0.25in}
    \epsfxsize=2.75in
    \epsffile{plots/final_r_kmu3_pipi0.epsi}
}
}
\caption{
$\rkekp$ and $\rkmukp$ results compared to the corresponding PDG value\,\cite{bib:PDG}.
\label{fig:results1}}
\end{figure}

\newpage

\end{document}